\documentclass{article}%
\usepackage{amsmath}%
\usepackage{amsfonts}%
\usepackage{amssymb}%
\usepackage{graphicx}

\usepackage{geometry}
\begin{document}

\title{Review of strategies \\ for a comprehensive simulation \\ in sputtering devices}
\author{Antonio A. Gentile \\ 
\small{\textit{Dipartimento di Ingegneria dell'Innovazione, University of Salento (Lecce 72100, Italy)}} \\
\small{\textit{\& MRS s.n.c. (Carmiano 73040, Italy) }} \\ 
\small{antonio.gentile@mrssnc.com} }
\date{\small{September 12$^{th}$, 2012}}
\maketitle

\begin{abstract}
The development of sputtering facilities, at the moment, is mainly pursued through experimental tests, or simply by expertise in the field, and relies much less on numerical simulation of the process environment. This leads to great efforts and empirically roughly optimized solutions: in fact, the simulation of these devices, at the state of art, is quite good in predicting the behavior of single steps of the overall deposition process, but it seems still ahead a full integration among the various tools already available for the phenomena involved in a sputter.  
We summarize here the various techniques and codes already available for problems of interest in sputtering facilities, and we try to outline the possible features of a comprehensive simulation framework, able to integrate the single paradigms in a full simulation, dealing with aspects going from the plasma environment up to the distribution and properties of the deposited film, not only on the surface of the substrate, but also on the walls of the process chamber. 
\end{abstract}

\section*{Introduction}

The simulation we are going to outline should deal with the specific case of Thin Film Sputtering processes, within the general framework of Physical Vapour Depositions (PVD). 
Basically, this kind of processes means operating in a ultra High-Vacuum (UHV) environment\footnote{a few mTorrs of pressure, filled up with neutral species like Ar}, with temperatures ranging from RT up to ab. $200^o$C, occasionally reached in certain parts of the system. A preferential direction of investigation will be devoted to \textsl{Magnetron Sputterers} (a strong magnetic field - ab. 200 Gauss - is superposed to the static/dynamic electric fields governing the behaviour of the plasma), since the magnetically aided ignition and maintenance of the plasma state has a significative improvement compared to non-magnetic technologies, and therefore is currently the standard in both research and industrial facilities \cite{Mat10}. Main advantages are in fact:
\newpage

\begin{itemize}
	\item reduction of the target-sample distance, and consequently higher efficiency;
	\item same ionization efficiencies at lower process gas pressure, which aids both the quality of the process and the deposition yield of sputtered atoms;
	\item no need to increase the voltage drop, therefore avoiding an increase in the electron mean free-path.
\end{itemize}

\smallskip
The long-term objective would be the full simulation of a sputter deposition process\footnote{A first attempt in this direction, with a success which is up to now very limited, has already been done with the development of the \textit{NEPTUNE Sputter} code \cite{Dep08} \cite{Miy00}}, which means the inclusion of the interdependence, due to their strong interactions, among the different phenomena concurring to the global process. In other words, a fully self-consistent, multi-physics simulation, where the calculated dynamics of each aspect is integrated with and directly affects all the others included in the simulation. A less ambitious approach would, in the first stage, deal with the various phenomena separately, neglecting or simplifying their mutual dependence: this would look much more like a collection and integration of state-of-art available simulation tools in the field (which would become the modules of the global approach), with the final scope of merging them into a framework, able to efficiently provide parameters, derived from one simulation module, as inputs to the modules directly related to it. \\ 
A straightforward novel application would be studying and predicting the interference of side-apparatus with the core sputtering facility. Specifically, this general approach could be applied to shields, simple or engineered ones, invasive sensors, non-standard materials employed inside a sputtering facility. These insertions rely currently more on experimental trials, rather than numerical simulation, which leads to empirically, roughly optimized solutions (given the extreme heterogeneity of the facilities available in this sector) and the need for new experimental tests - whenever changes to the facilities are made. 

\smallskip 
Expectations from the results of the simulation are more a hint about how the facility could react upon the insertions of non-standard features, and certainly not an \textit{ab-initio}, stand-alone calculation of optimizing parameters, given the assumption that 
\begin{quotation}
	\textit{treatment of a real plasma discharge in a reactor (is) not easily amenable to analytical or even numerical approaches, due to the variety of complex interactive phenomena} \cite{Har06}
\end{quotation}
Therefore, an interesting approach would be the progressive experimental validation of results from the simulation, and re-tuning of the parameters involved, until obtaining a good agreement with well-known, or easy-to-check, experimental data. Once this first result is established, one could move further to use simulation data as a realistic base for the optimization of currently available - as well as upcoming - technologies.  \\

\section{The Plasma source}

As stated before, this is probably the most difficult part to simulate\footnote{for a more general review of the approaches to the problem, which we synthesize here, we suggest \cite{Daw83}}. In fact it involves the dynamic simulation of motion and collisions (with appropriate cross-sections) between plasma particles (heavy ions and electrons) and background gas neutral atoms, with a particular focus on elastic and ionization collisions in the plasma environment. Parameters describing the plasmas can be highly variable, but gas pressures around $1 \, Pa$, with electron densities in the order of $10^{10}$ $cm^{-3}$, should be considered plausible. This requires an algorithm including a self-consistent simulation of the plasma charged particles motion, due to:
\begin{itemize}
	\item electric fields (mainly dc/rf, depending on the facility, with potential of $200-400$ V between cathode and anode; plus turbulent electric fields occurring in the \textit{electron trap} region \cite{Har06}, whose strength is a few V/cm) and 
  \item magnetic fields (in the case of Magnetrons: these can always be considered static, with an intensity of about $\cong 200 $ Gauss), 
\end{itemize}
including the mutual interactions between the sources of these two. Given the physics introduced above, it is a very used paradigm - in the general description of the plasma \cite{Ekp06} - the introduction of \textit{continuity} and \textit{Poisson} equations:
\begin{align}
	\frac{\partial n_{i(e)}}{\partial t} + \nabla \cdot \textbf{J$_{i(e)}$} = R_{i(e)} 
	\label{contin} \\
	\nabla^2 \phi = - \frac{q}{\epsilon_0} \left( n_i - n_e \right)
	\label{Poisson}
\end{align}
to match the behaviour of the ions (electrons) - in eq.\ref{contin} - each with its own density $n_{i(e)}$, with the determination of electrical potential $\phi$ - in eq.\ref{Poisson}. Notice that, given the ionization processes occurring in the plasma\footnote{Which will be described in detail in the next paragraph}, the continuity eqs. must embed a source term $R$. \\
The plasma environment has normally temperatures as low as possible, since hotter plasmas are more difficult to control, however, peak temperatures of the residual gas can reach also 900 K or more \cite{Jim07}. 

\subsection{The external fields}

A first step in the modeling of the overall 'plasma ignition' process relies upon an accurate calculation of the magnetic field generated by the permanent magnets inside the chamber and, in particular, in the proximity of the target's surface, exposed to the action of the plasma (the so called '\textit{sheath}' region). The flux density of the magnets is a known parameter, as well as their position, normally a few cm behind the cathode-target.   \\
It has been found in \cite{She90} that the particles' trajectories mostly affected by the action of the magnetic field are indeed the electrons' trajectories \footnote{since ions in the plasma are not significantly magnetized}. This is why the accuracy of the field calculation must be higher\footnote{this would mean a finer grid in the zone up to ab. 4 cm away from the cathode, or even the calculation to be performed in this zone only} where the density of the plasma electrons is maximal. The solution of this magnetostatic problem should pose no particular difficulty, and can be afforded i.e. with FEMLAB modeling, as it was done in \cite{Con05}. \\
An alternative approach is to introduce an effective potential ($\Psi$) picture:
\begin{equation}
	\Psi = \frac{(P_{\theta}-A_{\theta}rq)^2}{2mr^2} + q \phi
\end{equation}
where $q$ and $m$ are the electrons' charge and mass, $P_{\theta}$ its canonical momentum, whose position is described in a cylindrical system\footnote{particularly indicated for the case of a \textit{planar Magnetron}} of coordinates $(r,z,\theta)$, under the action of an electric potential $\phi$ and a magnetic vector potential $A_{\theta}$. In particular, the $z$ direction is the one along the target-substrate distance. In this way, one can make a rough description of the zones where most electrons are trapped, rather than, calculating the position and extension of the \textit{sheath} and \textit{pre-sheath}\footnote{the zone immediately further away from the cathode, where the tail of most energetic ions is generated} regions, using approximations:
\begin{itemize} 
	\item for $A_{\theta}$, given basic characteristics of the facility simulated, like the position of the target and the applied external magnetic field; 
	\item for $\phi$ in the sheath, instead, referring to theoretical models like Child's law \cite{She89}; 
	\item and finally $\phi$ in the pre-sheath is normally considered uniform \cite{She90}, or the small\footnote{If compared to the one occurring in the sheath region} potential drop in this zone can be accurately measured \cite{Gor90} and is in the order of a few V. 
\end{itemize}
Once $\Psi$ is known, the next step is the application of a Hamiltonian approach, like the one suggested in \cite{Wen88}, to derive the dynamic properties of the particles to be followed. \\ 
As a first approach to validate the simulating code, the surfaces of the chamber can be considered as perfectly conductive, grounded sheets, so to deal with an easier situation. A proposal to draw a more realistic picture in case of non-conductive insertions will be explained in the following. \\
In conclusion, the intensities of these external fields should be treated as known conditions (i.e. \textit{input parameters}) in the simulation, since the sources are known and controlled within a high degree of approximation. What is left to calculate (or directly measure) is the intensity of these fields within the real geometry of the equipment: the chamber where the process occurs, plus eventual insertions.

\subsection{Particles' Trajectories}

As emphasized in \cite{Goe90}, the general problem of particles' trajectories in a sputtering simulation can be split in at least two different frameworks, which will be treated in an almost independent way: electrons and ions. Neutrals will be included as static targets of the collisions occurring in the plasma, rather than as dynamic particles themselves.  

\smallskip
Electrons of interest in the plasma dynamics are the so called \textit{fast electrons}\footnote{i.e. those sufficiently energetic to ionize neutral atoms of the gas}. These are mostly generated in the sheath region, or in the proximity of the cathode \cite{Bun94}, by the combined action of the external fields and of the collisions with neutral particles. Some other are also the secondary electrons (also known as \textit{bulk electrons} or 'discharge electrons') generated as a side effect of the impact upon the target of the heavy positive ions (or to a minor extent, within the interaction with neutrals). It is possible to describe the phenomenon by an empirical yield $\gamma$, so that the secondary electrons current density $j_e$ at the target (position 0 on the z-axis):
\begin{equation}
	j_e (0) = -\gamma j_i (0)
	\label{2nd_yield}
\end{equation}
is related to the ions' current density $j_i$. \\ 
The motion of the electrons can be thus described starting from the integration of the equations of motion:
\begin{equation}
 \ddot{x} = \frac{q}{m} \left( \vec{E}+v\times\vec{B} \right)
 \label{e_motion}
\end{equation}
Once an efficient integrator for eq.(\ref{e_motion}) has been provided, one is able to track with high accuracy the location (i.e. orbits), velocity and kinetic energy of the electrons. In this way, one can improve the qualitative picture, given by the potential $\Psi$, analyzing as the first this \textit{collisionless motion} of the single electrons in the plasma\footnote{The calculation, for the case of a planar Magnetron, will indeed show that some of these orbits are confined near the cathode by the shape of $\Psi$, while others are unconfined and the corresponding electrons are able to 'escape' in direction of the anode, before releasing most of their energy in collisions. This collisionless description in Plasma Physics is a well established formalism also known as 'Vlasov-Maxwell equations'}.\\ 
Naturally, in order to derive realistic results, it is necessary to include the collisions among the electrons and the other species inside the sputtering facility, since the internal pressure is not so low to neglect them as in thermal evaporators. This will be the topic of the next subsection. 

\smallskip
In a sputter, ions are originated by some of the collisions occurring in the plasma\footnote{For further details see next subsection}, which means that these sites can be accurately predicted \cite{Goe90}, and lay in the electron trap region. Once the ionization has occurred, the calculation of ions' trajectories follows the same guidelines introduced for the electrons' case. \\ 
This means that the ions are subjected mainly to external fields (as already outlined, in particular to the dc/rf field, since the magnetic field has a reduced influence on ions, and could also be neglected). Apart from this major contribution in the $r$ and $z$ directions, experimental data \cite{Gor90}, have emphasized a noticeable amount of random energy in the $\theta$ direction. For this additional feature, normally turbulent electric fields\footnote{even if they are about 4 times smaller than the dc field already in the pre-sheath region}, or further collisions occurring in the plasma, are held responsible. 

\smallskip
Finally, notice that, within the drift-diffusion framework leading to eq.\ref{contin}, one can also describe globally the motion of both positive/negative charged particles in terms of a flux \textbf{J} \cite{Ekp06}, which is, including the action of the magnetic field for the specific case of a DC Magnetron:
\begin{align}
	\textbf{J}_{\alpha} = & \, (-) n_{\alpha} \left[ \mu_{\alpha}^{//} \textbf{E}^{//} 
	+ \mu_{\alpha}^{\small{\bot}} \textbf{E}^{\small \bot} + \mu_{\alpha}^d (\textbf{E}\times \textbf{h}) \right] \nonumber \\
	& - D_{\alpha}^{\small //} (\nabla n_{\alpha})^{//} - D_{\alpha}^{\small \bot} (\nabla n_{\alpha})^{\small \bot} + D_{\alpha}^d [\textbf{h} \times \nabla n_{\alpha}]
	\label{totalflux}
\end{align}
where \textbf{h} is the unit vector in \textbf{B} direction, and for each charged particle ($\alpha := \{ i,e\}$ ), we have defined its average rate of collisions with neutrals $\nu_{\alpha} (n) $, function of the density of the residual gas $n$, and therefore its effective mass $m^*_{\alpha}$, the components of the mobility:
\begin{equation}
	\mu^{//}_{\alpha} = \frac{q_{\alpha}}{m_{\alpha}^* \nu_{\alpha}} \, ; \,\,\,\,\,\, 
	\mu^d_{\alpha} = \frac{1}{B \left[ 1+ \left( \frac{m^*_{\alpha} \nu_{\alpha}}{m_{\alpha} \omega_{\alpha}} \right)^2 \right]} \, ; \,\,\,\,\,\,
	\mu^{\bot}_{\alpha} =  \frac{\mu^{//}_{\alpha}}{1+ \left( \frac{m^*_{\alpha} \nu_{\alpha}}{m_{\alpha} \omega_{\alpha}} \right)^2 }
	\label{mobility}
\end{equation}
and of the diffusion coefficients:
\begin{equation}
	D^{//}_{\alpha} = \frac{k_B T_{\alpha}}{m_{\alpha}^* \nu_{\alpha}} \, ; \,\,\,\,\,\, 
	D^d_{\alpha} = \frac{k_B T_{\alpha}}{q_{\alpha} B \left[ 1+ \left( \frac{m^*_{\alpha} \nu_{\alpha}}{m_{\alpha} \omega_{\alpha}} \right)^2 \right]} \, ; \,\,\,\,\,\,
	D^{\bot}_{\alpha} = \frac{D^{//}_{\alpha}}{1+ \left( \frac{m^*_{\alpha} \nu_{\alpha}}{m_{\alpha} \omega_{\alpha}} \right)^2 }
	\label{diffusion}
\end{equation}
for a certain cyclotron frequency $\omega_{\alpha} = q_{\alpha} B / m_{\alpha}$.

\subsection{Collision phenomena} \label{par_collision}

A variety of different kinds of collisions can occur in the plasma environment. A rough classification is given in the following. 
\begin{enumerate}
	\item \textit{Electrons VS Neutrals}. They can be of three main typologies: ionization, elastic scattering, excitation. The firsts are probably the most fundamental, since they generate and maintain the plasma itself: every time one occurs, the corresponding electron loses an energy equal to the ionization potential of the neutral specie, plus the kinetic energy of the (secondary) electron emitted. Elastic scattering produces a reduced energy loss of:
	\begin{equation} 
	\frac{\Delta K}{K} = \frac{4 m}{M} sin^2 (\alpha/2)
	\label{elastic_loss}
\end{equation}
Finally, there is the contribution from the excitation collisions. These last ones have been shown \cite{Car89} to be ab. 1\% of the total collisions, so are often neglected, along with Penning or double ionization phenomena. \\
It must be remembered that each kind of collision has its specific scattering cross section $d \sigma/d\Omega$, which varies with the energy of the incident electron. Detailed tabulated values of all three cross-sections are available in the literature \cite{Mcd64} and can be used in the simulation; interestingly, for highly energetic impacts\footnote{Which means, with kinetic energy $K>60$ eV, which often occurs in a typical Magnetron}, these cross sections are peaked at small angles whatever their type is. This means, most of the collisions occur without a substantial change in the canonical momentum of the electron\footnote{And this can be interpreted as a long permanence time for the electrons, compared to the average time between collisions, before they are scattered away from the trap}.
 \item \textit{Electrons VS Charged Particles} (or: \textit{Coulomb collisions}). Considering a typical Magnetron discharge, the frequency for these scattering events can be evaluated about five orders of magnitude less than the previous case \cite{Goe90}. This is why most of the models for the sputtering behaviour \textbf{neglect this kind of collisions}. 
 \item \textit{Ions VS Neutrals}. This kind of collisions alter significantly the trajectory of the involved ions, and therefore are important to understand the 'landing sites' of the ions on the target, and consequently the scattering process. In fact, these collisions (both elastic and charge exchange ones) are held co-responsible for the azimuthal randomness introduced before. To understand the role of these collisions, it has been estimated that an average ion created in the pre-sheath region of a Magnetron sputter has about 35\% probability to collide elastically with a neutral particle before reaching the target \cite{She90}. 
 \item \textit{Ions VS Ions}. These phenomena, which in principle occur with a significant probability, \textbf{can be neglected} for energetic reasons. In fact, in this process the energy transfer rate is much slower than the ion loss rate\footnote{As it can be verified from typical parameters of sputtering facilities, see i.e. \cite{Har06}}. 
\end{enumerate}
\smallskip
For this kind of simulations, Monte Carlo (MC) approaches are considered the standard in state-of-art literature, with a considerable variety of particular cases. Generally speaking, MC codes used in this case must be able to follow a large number of particles at a time\footnote{remember the densities are of about $10^{10}$ particles/cm$^3$ within a cylindrical zone, whose typical dimensions are 5 cm (for both radius and height)}, keeping track of their cinematic parameters and providing an accurate mechanism to simulate the recurrence of collisions, each with the adequate cross section. Usually, initial conditions for the particles should be unbiased in the code. \\
The first and rougher approach is to make use of the approximations cited above, plus an intuitive reduction of the number of particles involved in the simulation. This strategy relies on the consideration that, since charge/mass ratio remains the same, essential physics can be captured with a much smaller number of particles than those in a real plasma \cite{Daw83}. therefore, some simulations have been run with just 10$^3$ charged particles. \\
The usual way to introduce collisions in this numerical scheme is to divide the motion of the particles in time steps, and for each generate random numbers to be compared with the probability of a collision to occur, referring to tabulated values for the total cross sections available, for example, in \cite{deh79}. One afterwards discriminates between the various kinds of collisions by comparison among the different cross sections related to each case, again by generation of random numbers to decide whether the collision was i.e. elastic or ionizing. After that, the energy of the particles involved is varied in accordance to the rules stated above (and in \cite{Goe90}), and the motion after the scattering is altered in correspondence. Approaches of this kind has led to a noticeable variety of slightly different models, which go under the name of "Monte Carlo collision" (MCC) algorithms \cite{Bir91}. \\
A somehow more intelligent use of computer resources is based upon the so called "Particle-in-cell" (PIC) modeling. This type of alternative, competing algorithms deals with the problem in the 6D parameter space of the particles, subdivided by a grid in a large set of cells, which are 'efficiently sampled' by the particles themselves \cite{Bir91}. This means that we are making use of two main ideas:
\begin{itemize}
	\item if we ideally divide the phase space in cells, most of them are unoccupied most of the time, so that it is useless to include them in the calculation, complicating the resolution even in a Vlasov-Maxwell approach: therefore, each step of the simulation brings along a reduction in the number of cells which will be effectively treated in the algorithm at the next step;
	\item to perform a full simulation of the interactions among particles, it is not needed to calculate them all directly; rather than, it is possible to follow an averaged approach in 3 calculation steps for each time step: i) once the coordinates of the particles are known, they are used to calculate currents and charges deriving from the particles themselves; ii) these equations are integrated \textit{on the grid}, so to obtain values for the electromagnetic fields in each cell, which iii) will be used to extract the forces used in the next time step of the equations of motion integration.
\end{itemize}
These two ideas, though quite intuitive, have proven highly effective in reducing the computational time required for a full simulation\footnote{In example, it can be calculated that replacing the full naive 3D approach for 10$^8$ particles with a 64x64x64 cells simulation was able to reduce the computation time on a 10 Tflops cluster of about 6 orders of magnitude}. It is useful to remind that PIC algorithms are not necessarily collisionless, as they can introduce collision phenomena by the 'finite-size particles' scheme \cite{Daw83}. \\
A further improvement in the PIC approach is the so called 'implicit' version. In fact, within the traditional\footnote{Which in turn can be named the 'explicit' version} PIC algorithm described above, field equations need only the sources from the previous time cycle - and the equations of motions need only the fields from the previous time cycle - a simple procedure, whose limit is the severe stability conditions to be matched (see i.e. \cite{Lap10}). The power of the implicit approach, instead, relies in the implicit coupling, introduced between the discretization of the equations of motion for the particles\footnote{where \textit{p} is the particle index, $\theta$ is the discretized time-step, $n$ is the iteration step }
\begin{align}
	\textbf{x}_p^{n+1} &= \textbf{x}_p^{n} + \textbf{v}_p^{n+1} \Delta t  \nonumber \\
	\textbf{v}_p^{n+1} &= \textbf{v}_p^{n} + \frac{q_p \Delta t}{m_p} 
	\left( \textbf{E}_p^{n+\theta} (\textbf{x}_p^{n+1/2}) + \textbf{v}_p^{n+1/2} \times \textbf{B}_p^{n} (\textbf{x}_p^{n+1/2}) \right)
	\label{motion_discr}
\end{align}
and the discretization of the Maxwell equations for the fields:
\begin{align}
 \nabla \times \textbf{E}^{n+\theta} &= - \frac{1}{c} \frac{\textbf{B}^{n+1}-\textbf{B}^n}{\Delta t} \nonumber \\
 \nabla \times \textbf{B}^{n+\theta} &= \frac{1}{c} \frac{\textbf{E}^{n+1}-\textbf{E}^n}{\Delta t} + \frac{4\pi}{c} J^{n+\theta} \nonumber \\
 \nabla \cdot \textbf{E}^{n+\theta} &= -4 \pi \textbf{$\rho$}^{n+\theta} \nonumber \\
 \nabla \cdot \textbf{B}^{n+\theta} & = 0 
 \label{fields_discr}
\end{align}
Notice that the intermediate values for whatever variable, say $Q$, are computed as: \\ 
$Q^{n+\theta}=(1-\theta) Q^n + \theta Q^{n+1}$, where $\theta\in [1/2,1]$, since for $\theta < 1/2$ the algorithm is known to be unstable \cite{Bra82}. The main advantage deriving from these coupled equations\footnote{in fact, the presence of $\textbf{E}_p^{n+\theta}$ in eq.s \ref{motion_discr} requires the knowledge of the advanced electric field to calculate the particles' position, and viceversa, since the charges/densities rely on particles' sources in \ref{fields_discr}} is in the reduction to only a set of coupled fluid moment and field equations to be solved. Within a single time-step there is no more need of iterations to compute all of the parameters in the model. For a thorough description of the approximations introduced, to provide accurate solutions for the coupling, we refer to \cite{Lap10} and \cite{Bra82}. \\
We emphasize further that a successful implementation of these PIC schemes already exists and is provided by the CELESTE3D$^{\small \copyright}$ code\footnote{http://code.google.com/p/celeste/}, of which several releases are available. 

\smallskip
An interesting consideration to understand, for all the aspects of the plasma source described above, is: whether conditions are not too extreme to be treated within a fluid model, is there a chance to adopt multi-physics softwares\footnote{Such as i.e. the COMSOL$^{\small \textregistered}$ 'Plasma module' (hereafter the \textregistered will be omitted for brevity), eventually combined with other modules}? And by comparison of the results obtained by the software, with those from state-of-art MC techniques, does it exist a valid range of conditions, where such softwares can provide useful results with a considerable reduction of computation time and/or resources required? \\
The answer, up to the studies already available in the literature, can be considered carefully positive. For sure, in fact, a successful attempt in modeling (by a hybrid approach partially based on COMSOL modules) several parameters of a plasma discharge in a Magnetron device has been made in \cite{Jim07}, obtaining a good match with available experimental results. The assumptions made in the cited model are essentially:
\begin{itemize}
	\item the sources (ionization sites) in the plasma and the sampled particles' trajectories are calculated by a MC approach like the one suggested in \cite{She90}, under the same assumptions;
	\item gas heating effects\footnote{due to the energy lost by charged particles collisions with neutral particles of the residual gas} are taken into account in the simulation, and lead to rarefaction of the gas itself, which is modeled as ideal;
	\item data from the two points above are used as input parameters for the fluid-Poisson model, implemented in a COMSOL Multiphysics environment, solving the continuity and Poisson equations;
	\item the full 3D problem has been reduced, given symmetry properties\footnote{recalling the azimuthal non uniformity introduced in the previous paragraph, this should be considered a good approximation more than an exact hypothesis}, to a 2D mesh perpendicular to the cathode surface, and properly discretized. 
\end{itemize}
Additionally, the boundary conditions set for the equations (\ref{contin}) and (\ref{Poisson}) are:
\begin{align}
	\mbox{(for the electrons)    }  n_e &:= 0 \mbox{    (at cathode and walls)} \nonumber \\
  \mbox{(for the ions)    }  n_i &:= 0 \mbox{    (at walls, or defined by)    eq. (\ref{2nd_yield})    (at cathode)} \nonumber \\
    \phi &:= 0 \mbox{    (at walls, grounded and)    } \phi= - V_{appl} \mbox{    (at cathode)}
\end{align}
Notice that actually the only bound self-consistent with the MC calculation is for the cathodic ions' density: all the others rely on the assumption of a perfectly conducting, grounded surface of the walls. \\
The problem is afforded by the use of stationary 'Direct UMFPACK' solvers\footnote{http://www.cise.ufl.edu/research/sparse/umfpack/}, both linear and non-linear, to calculate the magnetic field inside the process chamber (with \textit{Magnetostatic} modules) and the COMSOL \textit{Chemical Engineering Module} for the solution of eqs. (\ref{contin}) and (\ref{Poisson}). The authors claim the results were in good agreement with experimentally available data from Langmuir probes, concerning the temperature of the residual gas, the electrons' densities and ions' flux, and finally the electrostatic potential in the chamber. \\
Results from the work above, by the way, should be considered carefully. A first major issue is that, in fact, the high ratio of the parallel magnetic field to cross field to electron mobility in a Magnetron is very high, and the fluid model involved in the \textit{Plasma module} code barely handles such extreme conditions. Such a module could thus be applied to the case of non-magnetically-aided facilities only, whereas for Magnetron-like environments hybrid approaches must be invoked. Another point to emphasize is the possible difficulty in dealing with RF sputtering (which up to now has been recalled quite a few), especially for the case of recent RF sputtering facilities employing Inductively Coupled Plasmas (ICP), because of a series of technical advantages\footnote{I.e. they avoid contact of metallic electrodes with the plasma and ionization mechanisms on isolating or conducting walls (the plasma is heated by the RF field induced by an external antenna), which leads to a high degree of plasma purity, lower pressures are possible, making a more directional technique available, and finally the absence of high voltage sheaths avoids potential co-sputtering from the walls.}. \\
ICP sputtering is simpler to simulate from some points of view: i) at pressures much lower than 1 mTorr, transport of ions can be considered collisionless, instead of diffusive; ii) because of the low plasma potential, ion bombardment of wall surfaces is reduced, and therefore the possibility of co-sputtering and charging effects\footnote{In fact, some ICP sputterers use even dielectric shields \cite{Har06}; a more detailed discussion of these phenomena will be presented below} is less significative. In any case, even if these aspects require less accuracy in the simulation, it turns out that the behaviour of the plasma itself can be tricky to simulate. In fact, in literature are reported cases of failure in the predictions about basic parameters of the plasma (like the electron density), based upon simulations run with the 'Plasma' and the 'RF' modules from COMSOL. A specific example is given by the low pressure regime of a jet ICP reactor, where important discrepancies were found between experimental results and the numerical simulation obtained for the plasma sources, based upon interaction between the electromagnetic fields and the process-gas flow \cite{Por11}.

\section{The sputtering process}
This process deals with the dynamics of energetic ions, which are accelerated towards the plane surface of the \textit{target} material. The impact with the target crystal generates \textit{recoil cascades} inside the lattice, displacing a certain number of atoms from their equilibrium sites. Those atoms on the target surface, which are not confined along the external direction, are hold on their sites from the \textit{Surface Binding Energy} ($E_b$), which is usually lower than the Displacement Energy of the bulk material \cite{Bun94}. Whenever these surface atoms acquire, from the cascades, an energy higher than $E_b$, they are extracted from the surface itself \cite{Mat10}, and this leads to their sputtering away from the target.  \\
Because of the physics of the process, a thin layer\footnote{Usually within the range 30-100 nm, from heavier to lighter ions, since these last ones tend to penetrate the target in deep \cite{Zie10}} of the target material is enough to perform an accurate simulation, reducing the time spent calculating cascades which will not contribute to sputtering. The most important value in describing the result of the sputtering process is the \textit{sputtering yield} $Y$, simply defined as:
\begin{equation}
	Y = \frac{\mbox{sputtered atoms}}{\mbox{incident ions}}
	\label{sputt-yield}
\end{equation}
where clearly, the higher the yield, the most effective is the ionic 'bombardment' for the deposition purposes. The yield itself can be obtained experimentally, or derived from semi-empirical models, like the Sigmund, Bohdanski, Yamamura or Wilhelm formulas \cite{Nak04}. Basically, there are two different approaches in solving the problem of simulating the sputtering yield and the energetic/angular distribution of the sputtered atoms. 

\smallskip 
A first choice is relying on experimental data, which are accurate and available for a considerable amount of different materials, in order to opportunely tune the parameters of semi-analytical models. In particular, the probability for a scattered atom to leave the surface of the target has a distribution $dJ/dE$ with the shape \cite{Cze89}:
\begin{equation}
	\frac{dJ}{dE}= C \frac{E}{(E+E_b)^2}
\end{equation}
rather than, as an input to the 'transport module' (see the next paragraph), it is possible to express the starting energy $E_0$ of each simulated atom as \cite{Pet99}:
\begin{equation}
	E_0 = \frac{\xi_E^{1/2} E_b}{(\kappa E_{bom} + E_b)/\kappa E_{bom} - \xi_E^{1/2}}
	\label{energy_distr}
\end{equation}
where $E_{bom}$ is the energy of the incident ion\footnote{which can be derived from further experimental data or as an input of the 'plasma source' environment}, $$ \kappa = 4 M_{ion} M_{atom} / (M_{ion} + M_{atom})^2 $$ while $\xi_E$ is a random number indicating the probability for an atom with energy $E_0$ to leave the target. A similar approach is used for the angular distributions, which are treated as respecting the Knudsen cosine law ($dJ/d \theta = D \cos \theta$), for the starting zenith angle $\theta_0$ \cite{Pet99}, and full randomness, for the azimuthal starting angle:
\begin{align}
   \theta_0 = \arcsin \xi_{\theta}
	\label{angle_zenit} \\
   \phi_0 = 2\pi \xi_{\phi}
	\label{angle_azimut}
\end{align}
where $\xi_{\theta}$ and $\xi_{\phi}$ are other random numbers. The starting trajectories sampled out in this way can be considered a rough approximation of the true input for the transport module, leading in any case to good predictions \cite{Pet99}.

\smallskip 
A more refined version of this approach involves instead the full treatment of the collisional dynamics in the solid, in order to retrieve the parameters above \cite{Mye91}, \cite{Nak04}. The standard is here again the adoption of MC algorithms, and in particular the \textit{TRIM} (TRansport of Ions in Matter) module of the \textit{SRIM} package\footnote{further details, papers and releases available under \underline{http://www.srim.org/}}. The core of the program is a detailed tracking of the incident ions and recoil processes, until their energy falls below a threshold indicating they are not anymore candidates for sputtering contributions \cite{Zie10}. Given that no assumptions are made about the lattice structure of the target, the code is reliable also in the case of amorphous material\footnote{rather than, polymeric materials of interest to us should not pose issues about the structure of the algorithm itself}. State-of-art implementation of the code makes use of the simpler but robust 'hard sphere' approximations (eq.\ref{hardsphere}) for the treatment of low-energy recoil events; more accurate theoretical tools are indeed available (like the 'Universal potential' of eq.\ref{unipot}, the 'Kr-C' or the recent 'ZBL electronic stopping' approaches \cite{Zie10}), and the last one in particular is used - in the TRIM package itself - for high energy events. \\
Basically, the TRIM code requires four input values for the simulation to be performed\footnote{Actually the code is provided with a library including over 28,000 different cases. However, since the package was calibrated for a specific case \cite{Nak04}, significant discrepancies with experimental results could be obtained in case no further calibration is performed}:
\begin{itemize}
	\item the lattice displacement energy and binding energy;
	\item the surface binding energy;
	\item the sputtering yield for normal incidence.
\end{itemize}
Moreover, the simulation allows to set a certain number of impacting ions, with a specific energetic and angular distribution: the effect of the angular incidence on the corresponding normal sputtering yield will be considered. Given these data, TRIM is able to calculate as output the sputtering yield dependency on ions' energy, the number of atoms displaced but not able to reach the surface of the target, and the angular distribution of the sputtered atoms \cite{Zie10}. However, it must be remembered that, if energetic plots have since a long time proved to provide good predictions - once accurately calibrated \cite{Mye91} - the angular plots (and therefore the direction of emitted atoms) seem to suffer from some intrinsic limit of the code \cite{Nak04}. A possible hint, about the reason for this discrepancy with experimental data, could be the roughening of the surface\footnote{Which is ignored from the TRIM code, and increases as the sputtering process continues.}, which both increases the yield - and this can be taken in account by minor adjustments - but significantly alters the angular distribution too, in a much less predictable way.

\smallskip
An alternative to the TRIM code is the implementation of dynamic MC modules (i.e. the so called \textit{SASA-sp}: the sputtering version of the dynamic-SASAMAL code), which are used in the 'NEPTUNE-Sputter' simulation environment \cite{Miy00}. In this case, the incident ion trajectory is seen as a set of straight-line segments of a length, which depends on the local atom density of the target material. This density is dynamically varied as the process goes on. At every vertex, a collision takes place with a certain specie of atom, according to the relative concentrations and scattering cross sections in the zone where the collision occurs. The electronic energy loss is subtracted from the kinetic energy of the incident ion, and a new motion/collision step in the algorithm is performed \cite{Miy00}. However, even if this code should have the enormous advantage of being part of an almost full simulation of sputtering facilities, it must be reported that it has been rarely cited - and therefore tested - up to now, in available literature\footnote{This could be due to the absence of an open-source code, which is the case for the TRIM code, or the reduced library of cases-of-study and optimizations}; among the few examples, see \cite{Dep08} (chap. 3).

\smallskip
Beside the traditional interest in the sputtering simulation from the target, an under-investigated problem is the role of the ions' impact against the walls or other side-apparatus (sometimes made of non-standard materials) inside the deposition chamber, which i.e. could lead to co-sputtering phenomena. In the specific case of amorphous dielectric structures used for shielding \cite{Mat10}, charging effects of the shields could also be possible. If so, it would be very interesting to understand the contribution from this charging to the global electromagnetic environment where the process occurs: this interdependence could make multi-physics approaches particularly adapt for these studies \footnote{Considering in example the COMSOL framework, it is not clear to us whereas any of its modules could be able to directly deal with the sputtering process problem. In fact, up to the knowledge of the author, applications of the COMSOL package in atomic extraction from surfaces have currently dealt with evaporative sources (\cite{Rou11}, or \cite{Con05}, where the COMSOL simulation is indeed limited to heat transfer effects) or CVD sources only (where the nature of the process is mainly chemical).}.

\section{The transport process}
The process of transport of the ejected particles, through the low pressure process gas, makes use of much of the formalism introduced for the plasma environment. An additional simplification allows to ignore the interaction with the applied electromagnetic fields: 
{in very good approximation}\footnote{For a typical clean metal or semiconductor surface, the percentage of charged emitted particles is only $\approx 10^{-4}$ of the total} \textit{the sputtered atoms are all neutral} \cite{Bun94}.  \\
Therefore, this step of the simulation essentially deals with the interaction among sputtered atoms, plasma particles and the background gas atoms. This scattering in gas phase can be modeled through different approaches, which can be roughly listed as follows, in order of increasing refinement:
\begin{enumerate}
	\item \textit{hard-sphere approximation}, which essentially relies on the simple definition of an average 'radius of interaction'   
	    $r_{\sigma}$, leading to a step potential: 
	\begin{equation} V = \begin{cases} 0 & \mbox{if  } |\textbf{r}_1- \textbf{r}_2|>r_{\sigma} \\ 
	                         V_{int} & \mbox{if  } |\textbf{r}_1- \textbf{r}_2|<r_{\sigma}  \end{cases} \label{hardsphere} \end{equation}
        this is especially used to model elastic collisions;
  \item \textit{Born-Mayer interatomic potentials}, which provide a range of interaction among the particles:
    \begin{equation} V =  \begin{cases}  Ae^{-B |\textbf{r}_1- \textbf{r}_2|}  & \mbox{    if   } r_{\sigma 1}<|\textbf{r}_1-      
      \textbf{r}_2|<r_{\sigma 2}  \\ 0 &  \mbox{    otherwise}  \end{cases} \end{equation}
	   a thorough investigation of plausible values of the parameters $A$, $B$, $r_{\sigma 1/2}$ has been given in \cite{Abr69};
	\item \textit{Lennard-Jones potentials}, of typical use in condensed matter physics, with the form:
	   \begin{equation} V =  4\lambda_{LJ} \left[ \left(\frac{\sigma}{r}\right)^{12}- \left(\frac{\sigma}{r}\right)^6 \right] \end{equation}
  \item a modified version \cite{Mye91} of the so called \textit{Universal interatomic potential}, which has the form\footnote{the van der Waals attractive contribution is introduced to fit the behavior at impact energies lower than 2 eV}:
      \begin{equation} V =  - 4\lambda_{LJ} \left(\frac{\sigma}{r}\right)^6 + \left(\frac{Z_1 Z_2 e^2}{r} \right) \chi_U (r)
                      \label{unipot}  \end{equation}
       where the 'universal screening function' $\chi_U$ has been used\footnote{it is semi-empirically defined as
	\[ \chi_U = 0.1818 e^{-3.2x} + 0.5099 e^{-0.9423x} + 0.2802 e^{-0.4028x} + 0.02817 e^{-0.2016x} \]
	  where the \textit{reduced distance} is $x= r (Z_1^{0.23} + Z_2^{0.23}) /0.89 \sigma$};
	\item the \textit{Firsov interatomic potential}, again based on a screen function\footnote{the so called Nikulin screen function, for more details see \cite{Pet99}} $\psi (r)$:
	      \begin{equation} V =  \left(\frac{Z_1 Z_2 e^2}{r} \right) \psi (r/x)
                        \end{equation}
\end{enumerate}
Whatever the particular inter-atomic potential chosen for the simulation, the scattering angle $\beta$ can be classically calculated from the integral definition:
\begin{equation} 
  \beta = -2 \pi \rho \int_{r_{min}}^{+\infty} {\frac{dr/r^2}{\sqrt{1-V(r)/E_{CM} - \rho^2/r^2}}}
 \label{scatter_angle}
\end{equation}
where $\rho$ is the \textit{collision parameter}. The calculation of the integral (\ref{scatter_angle}) reveals some issues which can be overcome by the application of proper strategies \cite{Mye91}. \\
Once accurate potentials have been introduced to describe the interactions of the scattered atoms with the residual gas, and the relative cross sections are known from tabulated or experimental data (see par. \ref{par_collision}), the simulation will be able to take into account both the thermalization and diffusion processes, along with the direct flow, involved in the transport of the sputtered species.   \\
A typical simulation scheme is the following. MC algorithms are used to introduce and calculate the effects (as described above) of scattering events in the direct flow of the emitted particles, at a rate dependent on the \textit{mean free path} of the particles at a certain pressure, as available from experimental data \cite{Pet99}. Once a particle has been thermalized by the collisions (e.g. its energy is within a $\sigma$ from the thermal energy $k_B T$), one can decide to keep tracking the particle by MC techniques, or rather spare computational resources switching to analytic diffusion equations. Interactions with charged particles and inelastic scattering are usually  neglected\footnote{Because ion and electron concentrations are small in comparison with the atom quantity (the range of ionization is $10^{-3}$), and average sputter atom energy (about 10 eV) is not enough to excite and ionize background gas atoms\cite{Bun94}}. The implementation of this transport step alone, given plausible description of the starting locations for the atoms, has proved able to provide results in good agreement with experimental data \cite{Mye91} \cite{Pet99}, especially concerning the thickness, angular and energetic distribution of the sputtered layer in the zone of the substrate. With the adoption of the most refined interactions, moreover, the accuracy was fully satisfying even for the case of multi-target depositions\footnote{With remarkably good results also for the final stoichiometry of the film \cite{Pet99}}. It is not fully clear, though, if the calculation of thicknesses on the chamber walls (e.g. far away from the substrate), explicitly shown in some results \cite{Mye91}, can be considered reliable or not. \\ 
An interesting point yet to investigate is whether a multiphysics 
approach\footnote{I.e. a 'Particle Tracing' module, example again taken from the COMSOL software framework}
would be able to embed one or several of the approaches above, with results at least comparable to those obtained from MC calculations, once the appropriate collision model has been defined in the software, .

\section{The deposition (Film growth)}
The effort in simulating the various steps of a sputtering process has, up to now, essentially ignored a properly contextual description of the deposition. In fact, since the transport of the sputtered species is essentially a physical process, and most of them are furthermore thermalized already before reaching the substrate \cite{Mye91}, no particular care about the surface chemistry is required, and the angular distribution is considered enough to understand how uniform the final film will be. Another point of view is the thorough analysis of mechanical/electric/thermal properties of the deposited film itself \cite{Bun94}, and therefore of the (micro)structures obtained on the substrate, neglecting any reference to the specific parameters of the deposition process. \\
However, we emphasize that it would be interesting to understand, even at a low-detail level, the interaction among the neutral transported species, on one side, and the substrate material (semiconductors, metals, plastics), where the accommodation on the final surface occurs. The main aim is to have a clue of the mechanical stress of the deposited material, depending on its thickness and deposition-rate, upon surfaces of different geometries or chemical-physical properties. An immediate, important application would be the optimization of the geometry and/or the material used for shielding solutions, in order to reduce the issue of flaking phenomena \cite{Mat10}. For these purposes, it is envisaged that a multi-physics approach 
%\footnote{i.e. the 'Structural Mechanics' module of COMSOL could be straightforwardly applied.} 
could be ideal in order to conjugate the effects of mechanical stress (deriving from the increasing, eventually non-uniform thickness of the deposited film), and the thermal stress (induced by the relatively hot plasma environment - by irradiation - and by the kinetic energy of the atoms impacting on the substrate surface). Given the nature of the phenomena to include in this last section, it should be considered the aspect of the global process less necessary to be integrated with the others. We claim, in fact, that an independent module specifically dealing with the film properties, according to data retrieved from the other three modules, would already be very useful to cope with this last aspect.

\section*{Conclusions}

In this work it has been given a thorough review of previously made attempts for the simulation of single phenomena in the field of sputtering processes. For each case, the methods available have been discussed, emphasizing their limits and perspectives. Whenever available, considerations about the effectiveness and the agreement with experimental data have been reported. \\
In addition, a complete framework embedding the various phenomena involved in a sputtering facility has been given, and the problem for a comprehensive interaction among available procedures is posed. Where plausible, adoption of features available in specific multi-physics software has been suggested as a possible strategy to simplify the simulation, compared to state-of-art numerical approaches. Given the current fragmentation of the approaches dedicated to the various aspects of the sputtering, this alternative strategy is considered worth to be investigated further.

\end{document}